\documentclass[12pt]{article}
\usepackage{amsmath}
\usepackage{amsfonts}
\usepackage{amsthm}

\newtheorem{fact}{Fact}[section]
\newtheorem{lemma}{Lemma}
\newtheorem{theorem}{Theorem}

\title{Coupling the Dirac and Einstein equations through geometry}
\author{jason hanson}

\begin{document}
\maketitle

\begin{abstract}
We show that the exterior algebra bundle over a curved spacetime can be used as framework in which both the Dirac and the Einstein equations can be obtained.  These equations and their coupling follow from the variational principle applied to a Lagrangian constructed from natural geometric invariants.  We also briefly indicate how other forces can potentially be incorporated within this geometric framework.
\end{abstract}

\section{Introduction}

The Dirac equation on Minkowski space ${\mathbb M}$ can be written in the form $i\gamma^\alpha\partial_\alpha\psi=m\psi$.  The gamma matrices $\gamma^\alpha$ give a four--dimensional representation of the Clifford algebra on ${\mathbb M}$.  These matrices act on a complex four--dimensional vector space $V$, and spinors $\psi$ are $V$--valued fields $\psi:{\mathbb M}\rightarrow V$.  Although $V$ and ${\mathbb M}$ are both four--dimensional spaces, they are fundamentally different.  Indeed if $\Lambda$ is a Lorentz transformation, then ${\bf v}'=\Lambda{\bf v}$ is the transformation rule for four--vectors in ${\mathbb M}$.  In contrast, the corresponding transformation rule for spinors in $V$ is $\psi'=S\psi$, where $S=\Sigma(\Lambda)$ is the image of $\Lambda$ under the spin representation $\Sigma$ of the Lorentz group constructed from the gamma matrices.

The standard approach to generalizing the Dirac equation to a curved spacetime $M$ is to assume that spinors are fields of a rank four complex vector bundle $E$ over $M$.  This bundle is assumed to be a Clifford module bundle with connection given by the spin connection $\nabla_\mu\psi=\partial_\mu\psi+\Gamma_\mu^{(s)}\psi$.  The spin connection matrix $\Gamma_\mu^{(s)}$ can be expressed in terms of the metric $\eta_{ab}$ and gamma matrices $\gamma_a$ for Minkowski space, as well as tetrads $e_\mu^a$ that determine the metric on $M$: $g_{\mu\nu}=e_\mu^ae_\nu^b\eta_{ab}$.  Indeed,
\begin{equation}\label{eq:spincon}
  \Gamma_\mu^{(s)}\doteq
  \tfrac{1}{8}([\gamma^\nu,\partial_\mu\gamma_\nu]
               -\Gamma_{\nu\mu}^\rho[\gamma^\nu,\gamma_\rho])
\end{equation}
where $\gamma_\mu=e_\mu^a\gamma_a$.  The Dirac equation on $M$ is then $i\gamma^\mu\nabla_\mu\psi=m\psi$.  See \cite{Pollock}.

A drawback of this approach is that the geometric and topological relationship between the bundle $E$ and spacetime $M$ is not specified a priori, so it is not clear how the Dirac equation should couple with gravity.  The main idea of this article is to use the exterior algebra bundle $\bigwedge_\ast M$ as a specific choice for $E$, thus completely specifying the relation between spinor space and spacetime.  However, the immediate price paid is that spinors are now sixteen--dimensional.  We will see that an even steeper price must be paid: the spin action constructed from the gamma matrices must be replaced by the action of the Lorentz group on $M$ when extended to $\bigwedge_\ast M$.  The spin action is still present, but it is relegated to the role of a discovered action.

The work presented here is related to the spacetime algebra formulation of the Dirac equation on Minkowski space by Hestenes \cite{Hestenes}.  The spacetime algebra is the Clifford algebra ${\it Cl}({\mathbb M})$ of Minkowski space, and spinors are fields of the real eight--dimensional subspace consisting of even degree elements.  Using the complex structure induced by the unit pseudoscalar of ${\it Cl}({\mathbb M})$, this subspace can also be identified as a four--dimensional complex vector space.  Under this identification, the spacetime algebra formulation is equivalent to the usual Dirac equation.  In contrast, the bundle $\bigwedge_\ast M$, when restricted to Minkowski space, can be identified with the entire Clifford algebra instead of a subspace.

{\it Article summary.}  We start off by reviewing the basic constructions used with the exterior algebra bundle, such as extending the metric, connection, and curvature.  We then indicate how these constructions transform under a change of coordinate basis, and write down some basic invariant quantities.  After comparing the extended Lorentz action with the spin action, we construct a Lagrangian from the geometric invariants and compute its variation with respect to (1) the spinor field, and (2) the metric.  We end by discussing how minimal coupling can be used to incorporate other forces.

Formulas given in this article that are not straightforward computations from the relevant definitions are verified in the appendix.

\section{Geometric framework}

Let $M$ denote spacetime.  That is, $M$ is a four--dimensional Lorentz manifold with metric $g$.  Throughout we let $x=x^\alpha$, for $\alpha=0,1,2,3$, denote local coordinates for $M$, and let ${\bf e}_\alpha\doteq\partial_\alpha=\partial/\partial x^\alpha$ be the corresponding basis for the fiber of the tangent bundle $T_\ast M$ of $M$ at $x$.  Except for signature of the metric, we adopt MTW \cite{MTW} conventions: $g_{\alpha\beta}\doteq g({\bf e}_\alpha,{\bf e}_\beta)$ are the components of $g$, and $g^{\alpha\beta}$ are the components of the metric inverse $g^{-1}$, so that $g_{\alpha\beta}\,g^{\beta\mu}=\delta_\alpha^\mu$.  Here and throughout the summation convention is used.  The canonical torsion--free metric--compatible connection on $T_\ast M$ is denoted by $\nabla$: $\nabla_\alpha{\bf e}_\beta=\Gamma_{\alpha\beta}^\mu{\bf e}_\mu$, where $\Gamma_{\alpha\beta}^\mu=g^{\mu\nu}\Gamma_{\nu\alpha\beta}$ and $\Gamma_{\nu\alpha\beta}=\tfrac{1}{2}(g_{\nu\alpha\beta}-g_{\alpha\beta\nu}+g_{\beta\nu\alpha})$, with $g_{\alpha\beta\nu}\doteq\partial_\nu g_{\alpha\beta}=\partial g_{\alpha\beta}/\partial x^\nu$.

{\it Sign conventions.} We will assume that the Lorentz metric has signature $(+,-,-,-)$, and that the Clifford algebra condition is $\gamma_\alpha\gamma_\beta+\gamma_\beta\gamma_\alpha=-2g_{\alpha\beta}$.  With these conventions, the Dirac equation in Minkowski space takes the form $\gamma^\alpha\partial_\alpha\psi=m\psi$.  Indeed for a plane wave $\psi=e^{-ip_\alpha x^\alpha}\psi_0$, the square of the Dirac operator is $\gamma^\alpha\gamma^\beta\partial_\alpha\partial_\beta\psi=(-i\gamma^\alpha p_\alpha)^2\psi=p^\alpha p_\alpha\psi=m^2\psi$.  So that the Dirac operator must have real eigenvalues.

\medskip
\noindent
{\bf Exterior algebra bundle.}
Let $\bigwedge_\ast M$ denote the complex exterior algebra bundle of $M$, which is formed by taking the exterior algebra of the fibers of the complexified tangent bundle of $M$.  For a multi--index $I$, the symbol
$${\bf e}_I\doteq{\bf e}_{\alpha_1}\wedge{\bf e}_{\alpha_2}
                  \wedge\cdots\wedge{\bf e}_{\alpha_k}
  \quad\text{with}\quad
  I=\alpha_1\alpha_2\cdots\alpha_k.
$$
denotes the exterior product of basis vectors from $T_\ast M$.  Set $|I|\doteq k$, and for $|I|=0$ we write ${\bf e}_I={\bf e}_\emptyset$ for the algebra unit.  The bundle $\bigwedge_\ast M$ is a complex vector bundle of rank sixteen, and has a local fiber basis given by elements ${\bf e}_I$ with $\alpha_1<\alpha_2<\cdots<\alpha_k$ and $0\leq k\leq 4$.  A spinor field $\psi$ on $\bigwedge_\ast M$ is a section, and we write $\psi=\psi^I{\bf e}_I$ for complex--valued functions $\psi^I=\psi^I(x)$.

Recall that the interior product $\iota_\alpha\psi$ of a basis vector ${\bf e}_\alpha$ and a spinor $\psi$ is defined by linear extension of the rules $\iota_\alpha{\bf e}_\beta=g_{\alpha\beta}{\bf e}_\emptyset$ and $\iota_\alpha({\bf e}_I\wedge{\bf e}_J)=(\iota_\alpha{\bf e}_I)\wedge{\bf e}_J+(-1)^{|I|}{\bf e}_I\wedge(\iota_\alpha{\bf e}_J)$.  Necessarily $\iota_\alpha{\bf e}_\emptyset=0$.

\medskip
\noindent
{\bf Extended metric.}
The metric $g$ on $T_\ast M$ can be extended to a Hermitian metric on $\bigwedge_\ast M$, which we denote as $\hat{g}$.  Setting $\hat{g}_{IJ}\doteq\hat{g}({\bf e}_I,{\bf e}_J)$, we have $\hat{g}_{IJ}=0$ if $|I|\neq|J|$, and
$$\hat{g}_{IJ}
  \doteq\det(g_{\alpha_i\beta_j})
  \quad\text{where}\quad
  I=\alpha_1\cdots\alpha_k
  \,\,\text{and}\,\,
  J=\beta_1\cdots\beta_k
$$
if $|I|=|J|=k$.  Note that $\hat{g}_{\emptyset\emptyset}=1$ and $\hat{g}_{\alpha\beta}=g_{\alpha\beta}$.  As a $16\times 16$ matrix, $\hat{g}$ is real symmetric.  For spinors $\psi,\phi$, we have $\hat{g}(\psi,\phi)=\psi^\dagger\hat{g}\phi=(\psi^I)^\ast\hat{g}_{IJ}\phi_J$.  The exterior and interior products are adjoint with respect to the extended metric: $\hat{g}({\bf e}_\alpha\wedge\psi,\phi)=\hat{g}(\psi,\iota_\alpha\phi)$.

\medskip
\noindent
{\bf Extended connection.}
The metric--compatible connection $\nabla$ on $T_\ast M$ can also be extended to $\bigwedge_\ast M$, which we denote by $\hat\nabla$, via the Leibniz rule:
$$\hat\nabla_\alpha(\phi\wedge\psi)
  =(\hat\nabla_\alpha\phi)\wedge\psi
   +\phi\wedge(\hat\nabla_\alpha\psi).
$$
Note that $\hat\nabla_\alpha{\bf e}_\emptyset=0$.  Define the extended Christoffel symbols $\hat\Gamma_{\alpha I}^J$ to be such that $\hat\nabla_\alpha{\bf e}_I=\hat\Gamma_{\alpha I}^J{\bf e}_J$.  We may write $\hat\nabla_\alpha\psi=\partial_\alpha\psi+\hat\Gamma_\alpha\psi$, where $\hat\Gamma_\alpha$ denotes the $16\times 16$ matrix with components $(\hat\Gamma_\alpha)_I^J=\hat\Gamma_{\alpha I}^J$.  The extended metric is compatible with the extended connection: $\hat{g}(\hat\nabla_\alpha\psi,\phi)+\hat{g}(\psi,\hat\nabla_\alpha\phi)=\partial_\alpha\hat{g}(\psi,\phi)$.  Equivalently in matrix form,
\begin{equation}\label{eq:metric-compat}
  \hat\Gamma_\alpha^\dagger\,\hat{g}+\hat{g}\,\hat\Gamma_\alpha
  =\partial_\alpha\hat{g}.
\end{equation}
Although the extended Christoffel symbols are real--valued: $\hat\Gamma_\alpha^\dagger=\hat\Gamma_\alpha^T$, we use the Hermitian conjugate symbol for consistency of notation.

\medskip
\noindent
{\bf Gamma matrices.}
For a basis element ${\bf e}_\alpha$, we define the gamma matrix $\gamma_\alpha$ to be the endomorphism of $\bigwedge_\ast M$ 
$$\gamma_\alpha\psi\doteq{\bf e}_\alpha\wedge\psi-\iota_\alpha\psi.$$
Using the definitions of the exterior and interior products, one verifies that
$$\gamma_\alpha\gamma_\beta+\gamma_\beta\gamma_\alpha=-2g_{\alpha\beta}
  \quad\text{and}\quad
  \gamma^\alpha\gamma^\beta+\gamma^\beta\gamma^\alpha
  =-2g^{\alpha\beta}
$$
where $\gamma^\alpha\doteq g^{\alpha\beta}\gamma_\beta$.  And from the fact that the exterior and interior products are adjoint with respect to the extended metric, we get
\begin{equation}\label{eq:gamma-xmetric}
  \gamma_\alpha^\dagger\,\hat{g}+\hat{g}\,\gamma_\alpha=0
\end{equation}
(note that $\gamma^\dagger=\gamma^T$).  Moreover, one can show that
\begin{equation}\label{eq:gamma-xconn}
  [\gamma_\alpha,\hat\Gamma_\beta]
  =\partial_\beta\gamma_\alpha-\Gamma_{\alpha\beta}^\epsilon\gamma_\epsilon
  \quad\text{and}\quad
  [\gamma^\alpha,\hat\Gamma_\beta]
  =\partial_\beta\gamma^\alpha+\Gamma_{\beta\epsilon}^\alpha\gamma^\epsilon.
\end{equation}

\medskip
\noindent
{\bf Extended curvature.}
Recall that the Riemann curvature operator $\Omega_{\alpha\beta}$ on $T_\ast M$ is given by $\Omega_{\alpha\beta}=\nabla_\alpha\nabla_\beta-\nabla_\beta\nabla_\alpha-\nabla_{[{\bf e}_\alpha,{\bf e}_\beta]}$.  Note that as we are assuming that the ${\bf e}_\alpha=\partial/\partial x^\alpha$ are coordinate frames, we have that $[{\bf e}_\alpha,{\bf e}_\beta]=0$.  By using the extended connection in place of the connection, we obtain the extended curvature operator $\hat\Omega_{\alpha\beta}$ for $\bigwedge_\ast M$.  In matrix form, we have
\begin{equation}\label{eq:xcurvature}
  \hat\Omega_{\alpha\beta}
  =\partial_\alpha\hat\Gamma_\beta
   -\partial_\beta\hat\Gamma_\alpha
   +[\hat\Gamma_\alpha,\hat\Gamma_\beta].
\end{equation}

\medskip
\noindent
{\bf Extended group action.}
Suppose that $A$ is a (local) transformation on $T_\ast M$, so that $A{\bf e}_\alpha=A_\alpha^\beta{\bf e}_\beta$.  We extend to a transformation on $\bigwedge_\ast M$:
\begin{equation}\label{eq:xaction}
  \hat{A}{\bf e}_I
  \doteq(A{\bf e}_{\alpha_1})\wedge\cdots\wedge(A{\bf e}_{\alpha_k})
  =A_{\alpha_1}^{\beta_1}\cdots A_{\alpha_k}^{\beta_k}
   {\bf e}_{\beta_1}\wedge\cdots\wedge{\bf e}_{\beta_k}
\end{equation}
for $I=\alpha_1\cdots\alpha_k\neq\emptyset$, and $\hat{A}{\bf e}_\emptyset\doteq{\bf e}_\emptyset$.  The expression on the right hand side of equation \eqref{eq:xaction} can be written as a sum in the basis of $\bigwedge_\ast M$.  By doing so, $\hat{A}$ can be written as a $16\times 16$ matrix with $\hat{A}{\bf e}_I=\hat{A}_I^J{\bf e}_J$.  Note that the matrix $\hat{A}$ contains the $4\times 4$ matrix $A$: $\hat{A}_\alpha^\beta=A_\alpha^\beta$.  Moreover if $A$ is invertible, then $(A^{-1})^{\hat{}}=\hat{A}^{-1}$.

\medskip

{\it Remarks.}
All constructions here are standard, although some of the formulas stated are not.  For the exterior algebra and the extended metric, see \cite{Sternberg}.  We are treating the exterior algebra as a Clifford algebra module, and for the gamma matrix formulation in this context, see \cite{Lawson}.  The extended group action is just the tensor product representation construction, as in \cite{Fulton}.  It should be noted that while the extended metric, connection, curvature, and group action all preserve the natural grading of $\bigwedge_\ast M$, the gamma matrices do not.

In a previous unpublished version of this article, I attempted to use the Clifford algebra bundle as the underlying geometric framework.  Similar constructions can be made, but extended metric compatibility, equation \eqref{eq:metric-compat}, does not hold without restrictions placed on the metric.

{\it Remark on notation.}
We have overloaded the meaning of the symbol $\,\hat{}\,$ for notational convenience at the expense of potential confusion.  The constructions for the extended metric $\hat{g}$, connection $\hat\Gamma$, curvature $\hat\Omega_{\alpha\beta}$, and group action $\hat{A}$ are all different.  The type of object that we apply $\,\hat{}\,$ to dictates the construction that should be used.

\section{Change of bases and invariants}

Let $B$ be a local change of basis for $T_\ast M$, so that we have the new fiber basis ${\bf e}_\alpha'=(B^{-1})_\alpha^\beta{\bf e}_\beta$.  Using \eqref{eq:xaction}, we extend $B$ to a change of basis for $\bigwedge_\ast M$, ${\bf e}_I'\doteq\hat{B}^{-1}{\bf e}_I=(\hat{B}^{-1})_I^J{\bf e}_J$.  We indicate how the geometric quantities of the previous section are affected by such a change of basis.

For a spinor field $\psi$ on $\bigwedge_\ast M$, we have $\psi^I{\bf e}_I=\psi=\psi'^I{\bf e}_I'$, so that $\psi'^I=\hat{B}_J^I\psi^J$ gives the transformation rule for fields on $\bigwedge_\ast M$.  In matrix form
\begin{equation}\label{eq:xfield}
  \psi'=\hat{B}\psi.
\end{equation}
The transformation rules for the extended metric, connection, and curvature matrices then follow by general principles:
\begin{align}
  \hat{g}'
    &=\hat{B}^{-T}\,\hat{g}\hat{B}^{-1}
     \label{eq:xmetric}\\
  \hat\Gamma_\alpha'
    &=(B^{-1})_\alpha^\beta(-\partial_\beta\hat{B}
                            +\hat{B}\hat\Gamma_\beta)\hat{B}^{-1}
     \label{eq:xchristoffel}\\
  \hat\Omega_{\alpha\beta}'
    &=(B^{-1})_\alpha^\rho(B^{-1})_\beta^\sigma
                           \hat{B}\,\hat\Omega_{\rho\sigma}\,\hat{B}^{-1}.
     \label{eq:curvature-trans}
\end{align}
Moreover the gamma matrices are linear transformations on $\bigwedge_\ast M$, so necessarily
\begin{equation}\label{eq:xgamma}
  \gamma_\alpha'
  =(B^{-1})_\alpha^\beta\,\hat{B}\,\gamma_\beta\,\hat{B}^{-1}
   \quad\text{and}\quad
  {\gamma'}^\alpha
  =B_\beta^\alpha\,\hat{B}\,\gamma^\beta\,\hat{B}^{-1}.
\end{equation}

Using equations \eqref{eq:xfield}, \eqref{eq:xchristoffel}, and \eqref{eq:xgamma} one computes that the Dirac operator on $\bigwedge_\ast M$
$$D\psi
  \doteq\gamma^\alpha\hat\nabla_\alpha\psi
  =\gamma^\alpha\partial_\alpha\psi+\gamma^\alpha\hat\Gamma_\alpha\psi.
$$
transforms like a spinor field:
\begin{equation}\label{eq:dirac}
  D'\psi'=\hat{B}D\psi.
\end{equation}
Consequently, the Dirac equation $D\psi=m\psi$ is invariant under an extended change of basis.

\begin{fact}\label{thm:invscalars}
The following scalars are invariant under an extended change of basis:
$$\psi^\dagger\,\hat{g}\,\psi,\quad
  \psi^\dagger\,\hat{g}\,D\psi,\quad
  {\rm tr}_k(\gamma^\alpha\gamma^\beta\hat\Omega_{\alpha\beta}).
$$
\end{fact}
\noindent
Here ${\rm tr}_k(A)$ denotes the $k$--th order trace of the $n\times n$ matrix $A$, and is defined by $\det(I+sA)=\sum_{k=0}^n s^k\,{\rm tr}_k(A)$.  In particular, ${\rm tr}_1(A)={\rm tr}(A)$, the usual trace of a matrix.  And ${\rm tr}_n(A)=\det(A)$, the determinant of $A$.

The above transformation rules apply to any invertible local transformation $B$, not just a Lorentz transformation.  In the latter case, we have the following.
\begin{fact}\label{fact:ext}
For a Lorentz transformation $\Lambda$, $\hat\Lambda^\dagger\hat{g}\hat\Lambda=\hat{g}$ and $\Lambda_\alpha^\beta\gamma_\beta=\hat\Lambda\gamma_\alpha\hat\Lambda^{-1}$.
\end{fact}
\noindent
That is, the extended metric is invariant under the extended Lorentz action.  And gamma matrices behave under a similarity transformation by the extended Lorentz action in the same way as expected for a spin action.

\section{Spin action}

The Lorentz algebra ${\it so}(g)$ consists of (local) transformations $L$ on $T_\ast M$ such that $L^Tg+gL=0$.  For $L\in{\it so}(g)$, the matrix $Lg^{-1}$ is anti--symmetric, and we obtain the (local) spin representation $\sigma$ of ${\it so}(g)$ by making the assignment
\begin{equation}\label{eq:spinrep}
  \sigma(L)
  \doteq-\tfrac{1}{8}L^{\alpha\beta}(\gamma_\alpha\gamma_\beta
                                     -\gamma_\beta\gamma_\alpha)
  =-\tfrac{1}{4}L^{\alpha\beta}\gamma_\alpha\gamma_\beta,
\end{equation}
where $L^{\alpha\beta}=L_\epsilon^\alpha g^{\epsilon\beta}$.  This defines a real Lie algebra representation.  Moreover, equation \eqref{eq:gamma-xmetric} and the antisymmetry of $L^{\alpha\beta}$ imply that $\sigma(L)$ is in the Lie algebra ${\it so}(\hat{g})$.  That is, $\sigma(L)^T\hat{g}+\hat{g}\sigma(L)=0$.

Exponentiation in ${\it so}(g)$ gives the proper orthochronous subgroup ${\it SO}_+(g)$ of $O(g)$: if $L\in{\it so}(g)$, then $\exp(L)$ is in ${\it SO}_+(g)$.  On the other hand, exponentiation of the spin Lie algebra representation gives the spin Lie group representation $\Sigma$ of ${\it SO}_+(g)$.  In particular if $\Lambda=\exp(L)$, then $\Sigma(\Lambda)=\exp(\sigma(L))$.  It should be noted that $\Sigma$ is actually a projective representation in that it is only defined up to sign.  Specifically, if $\exp(L_1)=\exp(L_2)$, then $\exp(\sigma(L_1))=\pm\exp(\sigma(L_2))$.

The Lie group ${\it SO}_+(g)$ thus acts locally on $\bigwedge_\ast M$: $\psi\mapsto\Sigma(\Lambda)\psi$.  We will call this the {\em spin action}.  Technically, as $\Sigma$ is only a projective representation, we are actually acting on the (fiber--wise) projectivization $P(\bigwedge_\ast M)$, where two elements in a fiber of $\bigwedge_\ast M$ are identified if one is a nonzero scalar multiple of the other.

\begin{fact}\label{fact:spin}
Let $\Lambda$ be a Lorentz transformation.  Then for $S\doteq\Sigma(\Lambda)$, $S^\dagger\hat{g}S=\hat{g}$ and $\Lambda_\alpha^\beta\gamma_\beta=S\gamma_\alpha S^{-1}$.
\end{fact}

\noindent
In particular, the extended metric is invariant under both the extended Lorentz and spin actions.

On the other hand, the spin action is not necessarily compatible with the extended connection.  Indeed, we have $\hat\nabla_\alpha S\psi=(\partial_\alpha S+[\hat\Gamma_\alpha,S])\psi+S\hat\nabla_\alpha\psi$.  So that $\hat\nabla_\alpha\psi'=S\hat\nabla_\alpha\psi$ if and only if $\partial_\alpha S=[S,\hat\Gamma_\alpha]$.  This is the case when $M={\mathbb M}$ is Minkowski space, as $\partial_\alpha S=0$ and $\hat\Gamma_\alpha=0$.  A consequence of this and Fact \ref{fact:spin} is that the Dirac equation $\gamma^\alpha\partial_\alpha\psi=m\psi$ is invariant under both the extended and spin actions on Minkowsi space.

Fields in $\bigwedge_\ast M$ are not spinors in the usual sense, as the transformation rule is $\psi'=\hat\Lambda\psi$, rather than $\psi'=S\psi$ with $S\doteq\Sigma(\Lambda)$, for a Lorentz transformation $\Lambda$.  While the two transformation rules lead to two different fields in the exterior algebra bundle, the question is whether or not they yield equivalent solutions to the Dirac equation.  In the usual formulation of the Dirac equation on Minkowski space, if we apply a similarity transformation to the gamma matrices, we obtain an equivalent Dirac equation.  So that if we use $\gamma_\alpha'=T\gamma_\alpha T^{-1}$ in place of $\gamma_\alpha$ in the Dirac equation, we get the solution $\psi'=T\psi$ instead of $\psi$.  These two solutions are equivalent, and they are in the same Lorentz frame if $\gamma_\alpha'=\gamma_\alpha$.  Observe that from Facts \ref{fact:ext} and \ref{fact:spin}, we have $(\hat\Lambda^{-1}S)\gamma_\alpha(\hat\Lambda^{-1} S)^{-1}=\gamma_\alpha$.  Using $T=\hat\Lambda^{-1}S$ as the similarity transformation, we see that $\hat\Lambda\psi$ and $S\psi$ are necessarily equivalent solutions of the Dirac equation on Minkowski space in the same Lorentz frame.

However, we cannot use the usual definition of Dirac adjoint as $\tilde\psi=\psi^\dagger\gamma^0$.  While $\psi^\dagger\gamma^0\psi$ is invariant under the spin action, it is not so under the extended Lorentz action.  There is a more natural choice for the Dirac adjoint on $\bigwedge_\ast M$, namely
\begin{equation}\label{eq:dadjoint}
  \tilde\psi\doteq\psi^\dagger\hat{g}.
\end{equation}
The quantity $\tilde\psi\psi$ is a real scalar that is invariant under both the extended Lorentz and spin actions.  With this definition, the current
$$j^\alpha\doteq-i\tilde\psi\gamma^\alpha\psi$$
transforms as a Lorentz four--vector under both actions.  Moreover, we have current conservation: ${j^\alpha}_{;\alpha}=\partial_\alpha j^\alpha+\Gamma_{\beta\alpha}^\beta j^\alpha=0$ for solutions of the Dirac equation.  The drawback of definition \eqref{eq:dadjoint} is that the scalar $\tilde\psi\psi$ is not necessarily positive.  Nevertheless, it is possible in Minkowski space to choose plane wave solutions for which it is.  Indeed, the space of solutions to $\gamma^\alpha\partial_\alpha\psi=m\psi$ is a vector space of dimension 8.  We can decompose this space into two summands of dimension 4, with $\tilde\psi\psi$ positive on one summand, and negative on the other.

\section{Variational formulas}

Let $\omega\doteq\sqrt{-{\it det}(g)}$ denote the spacetime density.  We may use the invariant scalars in Fact \ref{thm:invscalars} to form the invariant Lagrangian
\begin{equation}\label{eq:lagrangian}
  {\mathcal L}
  \doteq\int_M(L_D-mL_M-\kappa L_G)\,dV
\end{equation}
where $dV$ is the spacetime volume element, $\kappa$ is a constant, and
$$L_M\doteq\omega\,\psi^\dagger\,\hat{g}\,\psi
  \,\,\text{and}\,\,
  L_D\doteq\tfrac{1}{2}\omega(\psi^\dagger\,\hat{g}\,D\psi
                              + c.c.),
  \,\,\text{and}\,\,
  L_G\doteq\omega\,{\rm tr}(\gamma^\alpha\gamma^\beta\hat\Omega_{\alpha\beta}).
$$
We remark that we may also use $L_G=\omega R$, where $R$ is the scalar curvature of the non--extended connection on $T_\ast M$.  Both choices are equivalent, as we will see in section \ref{sec:gravity}.  However, the use of the extended curvature is more amenable to minimal coupling, as discussed in section \ref{sec:mcouple}.

\subsection{Field variation}

As $L_G$ does not depend on $\psi$, we only need to compute the variations of $L_M$ and $L_D$ with respect to $\psi^I$ and $\bar\psi^I$, where $\psi=\psi^I {\bf e}_I$ and $\bar\psi^I\doteq(\psi^I)^\ast$.  Write $L_M=\bar\psi^I\hat{g}_{IJ}\psi^J$.  Then $\delta L_M/\delta\bar\psi^I=\omega\hat{g}_{IJ}\psi^J$.  We will abbreviate this as
\begin{equation}\label{eq:mass-var}
  \frac{\delta L_M}{\delta\bar\psi}
  =\omega\,\hat{g}\psi.
\end{equation}
Similarly, variation with respect to $\psi^I$ leads to $\delta L_M/\delta\psi=\omega\,\psi^\dagger\hat{g}$.

We now compute the variation of $L_D$, with respect to $\psi$.  We will show that $\delta L_D/\delta\bar\psi^I=\omega\hat{g}_{IJ}(D\psi)^I$.  I.e.,
\begin{equation}\label{eq:dirac-var}
  \frac{\delta L_D}{\delta\bar\psi}
  =\omega\,\hat{g}D\psi.
\end{equation}
A similar computation will also yield $\delta L_D/\delta\psi=\omega\,(D\psi)^\dagger\hat{g}$.  Write
$$L_D=\tfrac{1}{2}\omega\bar\psi^I\hat{g}_{IJ}
        \gamma_K^{\alpha J}(\psi_\alpha^K+\hat\Gamma_{\alpha L}^K\psi^L)
      +c.c.
$$
where $\psi_\alpha\doteq\partial_\alpha\psi$.  It is well--known that $\partial_\alpha\omega=\Gamma_{\beta\alpha}^\beta\omega$.  Again suppressing the superscript $I$ in the variation with respect to $\bar\psi^I$, we find
\begin{align*}
  \frac{\delta L_D}{\delta\bar\psi}
  &=\frac{\partial L_D}{\partial\bar\psi}
    -\partial_\alpha\frac{\partial L_D}{\partial\bar\psi_\alpha}
   =\tfrac{1}{2}[\omega\hat{g}(D\psi)
       +\omega\hat\Gamma_\alpha^\dagger\gamma^{\alpha\dagger}\hat{g}\psi
       -\partial_\alpha(\omega\gamma^{\alpha\dagger}\hat{g}\psi)]\\
  &=\tfrac{1}{2}\omega[\hat{g}(D\psi)
      +\hat\Gamma_\alpha^\dagger\gamma^{\alpha\dagger}\hat{g}\psi
      -{\Gamma^\beta}_{\beta\alpha}\gamma^{\alpha\dagger}\hat{g}\psi
      -\gamma_{,\alpha}^{\alpha\dagger}\hat{g}\psi
      -\gamma^{\alpha\dagger}\hat{g}_{,\alpha}\psi
      -\gamma^{\alpha\dagger}\hat{g}\psi_\alpha]\\
  &=\tfrac{1}{2}\omega\bigl\{\hat{g}(D\psi)
      -(\gamma_{,\alpha}^\alpha-[\gamma^\alpha,\hat\Gamma_\alpha]
        +{\Gamma^\beta}_{\beta\alpha}\gamma^\alpha)^\dagger\hat{g}\psi\\
  &\hspace{72pt}
      -\gamma^{\alpha\dagger}(\hat{g}_{,\alpha}-\hat\Gamma_\alpha^\dagger\hat{g}
                              -\hat{g}\hat\Gamma_\alpha)\psi
      -\gamma^{\alpha\dagger}\hat{g}\hat\Gamma_\alpha\psi
      -\gamma^{\alpha\dagger}\hat{g}\psi_\alpha\bigr\}
\end{align*}
Using equations \eqref{eq:metric-compat} and \eqref{eq:gamma-xconn}, the second and third summands are zero.  And from equation \eqref{eq:gamma-xmetric}, the fourth and fifth summands combine to give $\hat{g}D\psi$.

The variation of ${\mathcal L}$ with respect to $\bar\psi$ therefore leads to the Dirac equation
\begin{equation}\label{eq:rdirac}
  D\psi=m\psi,
  \quad\text{where}\quad
  D\psi=\gamma^\alpha(\partial_\alpha\psi+\hat\Gamma_\alpha\psi).
\end{equation}
Variation with respect to $\psi$ yields the conjugated Dirac equation.

\subsection{Metric variation}
We now vary the individual summands in the Lagrangian \eqref{eq:lagrangian} with respect to the metric.  As expected, varying the gravity term $L_G$ will give the Einstein tensor.  Varying the mass and Dirac terms $L_M$ and $L_D$ will give a source term for the Einstein equation.

\subsubsection{Gravity term}\label{sec:gravity}

One computes that ${\rm tr}(\gamma^\alpha\gamma^\beta\hat\Omega_{\alpha\beta})=8R$, where $R$ is the scalar curvature of $M$.  Thus,
$$L_G=8\omega R.$$
Standard results from general relativity (see \cite{MTW}) then imply that
\begin{equation}\label{eq:lgvar}
  \frac{\delta L_G}{\delta g_{\alpha\beta}}
  =8\omega\,G^{\alpha\beta}
\end{equation}
with $G^{\alpha\beta}=R^{\alpha\beta}-\tfrac{1}{2}g^{\alpha\beta}R$ the Einstein tensor, and $R_{\alpha\beta}$ the Ricci curvature tensor.

\subsubsection{Mass and Dirac terms}

Recall that $\partial\omega/\partial g_{\alpha\beta}=\tfrac{1}{2}g^{\alpha\beta}\omega$.  As $L_M=\omega\psi^\dagger\hat{g}\psi$, we consequently have
$$\frac{\delta L_M}{\delta g_{\alpha\beta}}
  =\frac{\partial L_M}{\partial g_{\alpha\beta}}
  =\omega\psi^\dagger A^{\alpha\beta}\psi
  \quad\text{where}\quad
  A^{\alpha\beta}
  \doteq\tfrac{1}{2}g^{\alpha\beta}\hat{g}
        +\frac{\partial\hat{g}}{\partial g_{\alpha\beta}}.
$$
Note that the matrices $A^{\alpha\beta}$ are real and symmetric, so that $(A^{\alpha\beta})^\dagger=A^{\alpha\beta}$.

Now consider $L_D=\tfrac{1}{2}\omega\psi^\dagger\hat{g}\gamma^\rho(\partial_\rho\psi+\hat\Gamma_\rho\psi)+c.c.$, which is a function of the metric and its first order derivatives.  One computes that
$$\frac{\delta L_D}{\delta g_{\alpha\beta}}
  =\frac{\partial L_D}{\partial g_{\alpha\beta}}
   -\partial_\epsilon\frac{\partial L_D}{\partial g_{\alpha\beta\epsilon}}
  =\tfrac{1}{2}\omega\left(
     \psi^\dagger A^{\alpha\beta}D\psi
     +\psi^\dagger P^{\alpha\beta}\psi
     +\psi^\dagger Q^{\alpha\beta\epsilon}\hat\nabla_\epsilon\psi
     +c.c.\right)
$$
with $A^{\alpha\beta}$ as above, and
\begin{align*}
  P^{\alpha\beta}
  &\doteq
   \hat{g}\gamma^\nu\frac{\partial\hat\Gamma_\nu}{\partial g_{\alpha\beta}}
   -\Gamma_{\mu\epsilon}^\mu\hat{g}\gamma^\nu
      \frac{\partial\hat\Gamma_\nu}{\partial g_{\alpha\beta\epsilon}}
   -\hat{g}_{,\epsilon}\gamma^\nu
      \frac{\partial\hat\Gamma_\nu}{\partial g_{\alpha\beta\epsilon}}
   -\hat{g}\gamma_{,\epsilon}^\nu
      \frac{\partial\hat\Gamma_\nu}{\partial g_{\alpha\beta\epsilon}}
   -\hat{g}\gamma^\nu
      \partial_\epsilon\frac{\partial\hat\Gamma_\nu}{\partial g_{\alpha\beta\epsilon}}\\
   &\quad\quad
   +\hat\Gamma_\epsilon^\dagger\hat{g}\gamma^\nu
      \frac{\partial\hat\Gamma_\nu}{\partial g_{\alpha\beta\epsilon}}
   +\hat{g}\gamma^\nu\frac{\partial\hat\Gamma_\nu}
                          {\partial g_{\alpha\beta\epsilon}}
       \hat\Gamma_\epsilon\\
  Q^{\alpha\beta\epsilon}
  &\doteq
    \hat{g}\frac{\partial\gamma^\epsilon}{\partial g_{\alpha\beta}}
    -\frac{\partial\hat\Gamma_\nu^\dagger}{\partial g_{\alpha\beta\epsilon}}
       \gamma^{\nu\dagger}\hat{g}
    -\hat{g}\gamma^\nu
       \frac{\partial\hat\Gamma_\nu}{\partial g_{\alpha\beta\epsilon}}.
\end{align*}
If we use equations \eqref{eq:metric-compat}, \eqref{eq:gamma-xmetric}, and \eqref{eq:gamma-xconn}, and the fact that $\gamma^\nu$ and $\hat{g}$ do not depend on the first derivatives of the metric, we can rewrite the previous equations in the form
\begin{equation}\label{eq:var1}
  \begin{aligned}
    P^{\alpha\beta}&=\hat{g}\gamma^\nu P_\nu^{\alpha\beta}\\
    P_\nu^{\alpha\beta}
    &=\frac{\partial\hat\Gamma_\nu}{\partial g_{\alpha\beta}}
      +\Gamma_{\epsilon\nu}^\mu
       \frac{\partial\hat\Gamma_\mu}{\partial g_{\alpha\beta\epsilon}}
      -\Gamma_{\mu\epsilon}^\mu
       \frac{\partial\hat\Gamma_\nu}{\partial g_{\alpha\beta\epsilon}}
      -\partial_\epsilon\frac{\partial\hat\Gamma_\nu}{\partial g_{\alpha\beta\epsilon}}
      +\bigl[\frac{\partial\hat\Gamma_\nu}
                  {\partial g_{\alpha\beta\epsilon}},
             \hat\Gamma_\epsilon\bigr]
  \end{aligned}
\end{equation}
\begin{equation}\label{eq:var2}
\begin{aligned}
  Q^{\alpha\beta\epsilon}&=\hat{g}T^{\alpha\beta\epsilon}\\
  T^{\alpha\beta\epsilon}&=
    \frac{\partial\gamma^\epsilon}{\partial g_{\alpha\beta}}
    +\frac{\partial}{\partial g_{\alpha\beta\epsilon}}
        \bigl(\hat{g}^{-1}\hat{g}_{,\nu}\gamma^\nu
              -\{\gamma^\nu,\hat\Gamma_\nu\}\bigr)
\end{aligned}
\end{equation}
where $\{\cdot,\cdot\}$ denotes the anti--commutator.

Observe that $P_\nu^{\alpha\beta}$ preserves the grading of $\bigwedge_\ast M$ and satisfies the Leibniz rule: $P_\nu^{\alpha\beta}(\psi\wedge\phi)=(P_\nu^{\alpha\beta}\psi)\wedge\phi+\psi\wedge(P_\nu^{\alpha\beta}\phi)$.  One computes $P_\nu^{\alpha\beta}{\bf e}_\rho=0$.  It follows that $P_\nu^{\alpha\beta}$ is trivial, whence
$$P^{\alpha\beta}=0.$$
On the other hand, the matrices $Q^{\alpha\beta\epsilon}$ are not trivial.  Indeed, we find that the associated matrices $T^{\alpha\beta\epsilon}$ can be computed from the rules
\begin{equation}\label{eq:Tmatrix}
\begin{aligned}
  T^{\alpha\beta\epsilon}\psi
  &=\gamma^\nu T_\nu^{\alpha\beta\epsilon}\psi
    +\tfrac{1}{2}(g^{\alpha\beta}\gamma^\epsilon
      -g^{\beta\epsilon}\gamma^\alpha
      -g^{\alpha\epsilon}\gamma^\beta)\psi\\
  T_\nu^{\alpha\beta\epsilon}(\psi\wedge\phi)
  &=(T_\nu^{\alpha\beta\epsilon}\psi)\wedge\phi
    +\psi\wedge(T_\nu^{\alpha\beta\epsilon}\phi)\\
  T_\nu^{\alpha\beta\epsilon}{\bf e}_\rho
  &=\tfrac{1}{2}[\delta_\nu^\alpha(g^{\epsilon\mu}\delta_\rho^\beta
                       -g^{\beta\mu}\delta_\rho^\epsilon)
     +\delta_\nu^\beta(g^{\epsilon\mu}\delta_\rho^\alpha
                       -g^{\alpha\mu}\delta_\rho^\epsilon)]
     {\bf e}_\mu.
\end{aligned}
\end{equation}

We remark that the matrices $T^{\alpha\beta\epsilon}$ defined in equation \eqref{eq:Tmatrix} are divergence--free in the first two indices.  Explicitly, we have
$${T^{\alpha\beta\epsilon}}_{;\beta}
  \doteq
   {T^{\alpha\beta\epsilon}}_{,\beta}
   +[\hat\Gamma_\beta,T^{\alpha\beta\epsilon}]
   +\Gamma_{\lambda\beta}^\alpha T^{\lambda\beta\epsilon}
   +\Gamma_{\lambda\beta}^\beta T^{\alpha\lambda\epsilon}
   +\Gamma_{\lambda\beta}^\epsilon T^{\alpha\beta\lambda}
  =0
$$
and ${T^{\alpha\beta\epsilon}}_{;\alpha}=0$ (note that $T^{\alpha\beta\epsilon}=T^{\beta\alpha\epsilon}$).

\subsubsection{Coupled Einstein equation}

Taking the above computations together, we find that the variation of ${\mathcal L}$ with respect to $g_{\alpha\beta}$ leads to the Einstein equation with source term:
$$G^{\alpha\beta}
  =\tfrac{1}{16\kappa}
   [\psi^\dagger A^{\alpha\beta}(D\psi-m\psi)
    +\psi^\dagger\hat{g}T^{\alpha\beta\epsilon}\hat\nabla_\epsilon\psi
    +c.c.].
$$
In particular, if $\psi$ satisfies the Dirac equation $D\psi=m\psi$, then we have
\begin{equation}\label{eq:einstein}
  G_{\alpha\beta}
  =\tfrac{1}{16\kappa}
   (\psi^\dagger\hat{g}{T_{\alpha\beta}}^\epsilon\,\hat\nabla_\epsilon\psi
    +c.c.)
\end{equation}
after lowering indices.

\section{Minimal coupling}\label{sec:mcouple}

To incorporate forces other than gravity into our framework, we use minimal coupling.  That is, we replace the extended connection $\hat\Gamma_\alpha$ in equation \eqref{eq:rdirac} with the {\em total connection}
\begin{equation}\label{eq:totcon}
  C_\alpha=\hat\Gamma_\alpha+\theta_\alpha.
\end{equation}
Here the $\theta_\alpha$ are $16\times 16$ matrices that correspond to the additional force.  However, these matrices must satisfy some constraints in order for the field variation of the sum of Lagrangian densities $L_D-mL_M$ to yield the minimally coupled Dirac equation.  That is, if we use the total connection $C_\alpha$ in place of the extended connection, then we should still arrive at equation \eqref{eq:dirac-var} when we vary $L_D$ with respect to the field $\psi$.  By examining the computation that follows equation \eqref{eq:dirac-var}, we see that the two conditions
$$C_\alpha^\dagger\hat{g}+\hat{g}C_\alpha
  =\partial_\alpha\hat{g}
  \quad\text{and}\quad
  [\gamma^\alpha,C_\beta]
  =\partial_\beta\gamma^\alpha+\Gamma_{\beta\epsilon}^\alpha\gamma^\epsilon
$$
are sufficient to do this.  On the other hand, the extended connection already satisfies equations \eqref{eq:metric-compat} and \eqref{eq:gamma-xconn}.  So the matrices $\theta_\alpha$ must satisfy
\begin{equation}\label{eq:force}
  \theta_\alpha^\dagger\hat{g}+\hat{g}\theta_\alpha=0
  \quad\text{and}\quad
  [\gamma^\alpha,\theta_\beta]=0.
\end{equation}

Observe that the diagonal matrices $\theta_\alpha=iA_\alpha$, with $A_\alpha$ real--valued scalars, satisfy the two conditions in equation \eqref{eq:force}.  So that electromagnetism is readily incorporated, provided one chooses an appropriate Lagrangian.  E.g., either equation \eqref{eq:LtrFsq} or \eqref{eq:Ltr2F}, discussed below.  Other distinct types of matrices that satisfy equation \eqref{eq:force}, and the corresponding physical forces allowed, will be examined in a separate article.

We remark that it is not clear how to incorporate non--gravitational forces into the Lagrangian, equation \eqref{eq:lagrangian}.  The ``minimal'' choice would be to set
$$L_G=\omega\,{\rm tr}(\gamma^\alpha\gamma^\beta F_{\alpha\beta})$$
where $F_{\alpha\beta}$ is the curvature of the total connection in equation \eqref{eq:totcon}.  However, in flat spacetime where the extended connection is trivial, the fact that $\theta_\alpha$ (and hence $F_{\alpha\beta}$) commutes with the gamma matrices will imply that $L_G=0$.  To avoid this, we can make the choice
\begin{equation}\label{eq:LtrFsq}
  L_G=\omega\,{\rm tr}(F_{\alpha\beta}
                       \,F^{\alpha\beta}).
\end{equation}
Another choice is
\begin{equation}\label{eq:Ltr2F}
  L_G=\omega\,{\rm tr}_2(\gamma^\alpha\gamma^\beta
                         F_{\alpha\beta})
\end{equation}
where ${\rm tr}_2$ is the 2--trace mentioned after Fact \ref{thm:invscalars}.  In the case of a flat spacetime, these two choices are actually equal.  Also possible, but even more computationally formidable, is
\begin{align*}
  L_G
  &=\omega\,\det({\it id}+\tau\gamma^\alpha\gamma^\beta F_{\alpha\beta})\\
  &=\omega
    +\omega\tau\,{\rm tr}(\gamma^\alpha\gamma^\beta F_{\alpha\beta})
    +\omega\tau^2\,{\rm tr}_2(\gamma^\alpha\gamma^\beta F_{\alpha\beta})
    +\cdots
    +\omega\tau^{16}\,\det(\gamma^\alpha\gamma^\beta F_{\alpha\beta})
\end{align*}
where $\tau$ is a constant.

\section{Concluding remarks}

The key to our geometric framework that allows us to couple the Dirac and Einstein equations is taking spinors to be sections of the exterior bundle over spacetime.  However, this implies that spinors must transform via the induced non--spin representation of the Lorentz group.  This is different from the standard formulation of the Dirac equation, wherein spinors transform via a spin representation.  Nonetheless, we can recover the spin action as a discovered action.  And in flat spacetime, the Dirac equation in the exterior bundle formulation is invariant under both spin and non--spin actions, so reproduces the standard formulation of the Dirac equation --- albeit with sixteen dimensional spinors instead of the usual four.

On the other hand, abandoning the requirement that spinors transform under the spin action might seem too steep of a price to pay.  After all, the spin action assumption is bound to the terminology used to describe a Fermion: {\em spinor} for the particle wave function, and {\em spin up/down} for the particle states.  However, particle states arise from using a wave function that is a field of more than one dimension, and not necessarily from the spin representation itself.  So perhaps the price is only a change in interpretation.


\appendix
\section*{Appendix: computations}

Here we sketch the verification of some of the formulas stated in this article that are not necessarily straightforward computations.

\begin{lemma}\label{lem:g-wedge}
$\hat{g}({\bf e}_\alpha\wedge\psi,{\bf e}_\beta\wedge\phi)=g_{\alpha\beta}\,\hat{g}(\psi,\phi)-\hat{g}(\iota_\beta\psi,\iota_\alpha\phi)$
\end{lemma}

\begin{proof}
Apply the identity $\hat{g}({\bf e}_\alpha\wedge\psi,\phi)=\hat{g}(\psi,\iota_\alpha\phi)$ twice.
\end{proof}

\begin{lemma}\label{lem:con-int}
$[\hat\nabla_\beta,\iota_\alpha]=\Gamma_{\alpha\beta}^\epsilon\,\iota_\epsilon$
\end{lemma}

\begin{proof}
First compute $[\hat\nabla_\beta,\iota_\alpha]({\bf e}_\rho\wedge\psi)
 =\Gamma_{\rho\alpha\beta}\,\psi-{\bf e}_\rho\wedge[\hat\nabla_\beta,\iota_\alpha]\psi$.  Now use induction on the grading degree, along with facts (i) $[\hat\nabla_\beta,\iota_\alpha]{\bf e}_\emptyset=0$, (ii) $[\hat\nabla_\beta,\iota_\alpha]\psi=\psi^I[\hat\nabla_\beta,\iota_\alpha]{\bf e}_I$, and (iii) $\Gamma_{\alpha\beta}^\epsilon\,\iota_\epsilon({\bf e}_\rho\wedge\phi)=\Gamma_{\rho\alpha\beta}\,\phi-{\bf e}_\rho\wedge\Gamma_{\alpha\beta}^\epsilon\,\iota_\epsilon\phi$.
\end{proof}

\begin{theorem}
$[\gamma_\alpha,\hat\Gamma_\beta]=\partial_\beta\gamma_\alpha-\Gamma_{\alpha\beta}^\epsilon\gamma_\epsilon$
\end{theorem}

\begin{proof}
This follows by computing $\nabla_\beta\gamma_\alpha\psi$ in two different ways.  First, viewing $\gamma_\alpha$ as a matrix, we find $\hat\nabla_\beta\gamma_\alpha\psi=(\partial_\beta\gamma_\alpha)\psi-[\gamma_\alpha,\Gamma_\beta]\psi+\gamma_\alpha\hat\nabla_\beta\psi$.  Second, using the definition $\gamma_\alpha\psi={\bf e}_\alpha\wedge\psi-\iota_\alpha\psi$ and lemma \ref{lem:con-int}, we obtain $\hat\nabla_\beta\gamma_\alpha\psi=\Gamma_{\alpha\beta}^\epsilon\,\gamma_\epsilon\psi+\gamma_\alpha\hat\nabla_\beta\psi$.
\end{proof}

\begin{theorem}
$\hat{g}(\hat\nabla_\beta\psi,\phi)+\hat{g}(\psi,\hat\nabla_\beta\phi)=\partial_\beta\hat{g}(\psi,\phi)$
\end{theorem}

\begin{proof}
It suffices to assume that $\psi$ and $\phi$ are homogeneous of the same grading degree.  The statement is true in degrees 0 (trivially) and 1 (by definition).  Use lemmas \ref{lem:g-wedge} and \ref{lem:con-int} to compute
\begin{align*}
  &\hat{g}\bigl(\hat\nabla_\beta({\bf e}_\rho\wedge\psi),
                {\bf e}_\sigma\wedge\phi\bigr)
   +\hat{g}\bigl({\bf e}_\rho\wedge\psi,
                 \hat\nabla_\beta({\bf e}_\sigma\wedge\phi)\bigr)\\
  &\quad\quad
   =g_{\rho\sigma\beta}\,\hat{g}(\psi,\phi)
    +g_{\rho\sigma}\,[\hat{g}(\hat\nabla_\beta\psi,\phi)
                      +\hat{g}(\psi,\nabla_\beta\phi)]\\
  &\quad\quad
   \quad\quad
    -[\hat{g}(\iota_\sigma\psi,\hat\nabla_\beta\iota_\rho\phi)
      +\hat{g}(\hat\nabla_\beta\iota_\sigma\psi,\iota_\rho\phi)].
\end{align*}
Induction on the grading degree implies that the right hand side of this equation is equal to
$$g_{\rho\sigma\beta}\,\hat{g}(\psi,\phi)
  +g_{\rho\sigma}\,\partial_\beta\hat{g}(\psi,\phi)
  -\partial_\beta\hat{g}(\iota_\sigma\psi,\iota_\rho\phi)
  =\partial_\beta\hat{g}({\bf e}_\rho\wedge\psi,{\bf e}_\sigma\wedge\phi),
$$
courtesy of lemma \ref{lem:g-wedge} again.
\end{proof}

\begin{theorem} For any Lorentz transformation $\Lambda$, $\hat\Lambda\gamma_\alpha\hat\Lambda^{-1}=\Lambda_\alpha^\beta\gamma_\beta$
\end{theorem}

\begin{proof}
First show that $\hat\Lambda\iota_\alpha\hat\Lambda^{-1}=\Lambda_\alpha^\beta\iota_\beta$.  This is obtained by the computation $(\hat\Lambda\iota_\alpha\hat\Lambda^{-1}-\Lambda_\alpha^\beta\iota_\beta)({\bf e}_\rho\wedge\psi)=-{\bf e}_\rho\wedge(\hat\Lambda\iota_\alpha\hat\Lambda^{-1}-\Lambda_\alpha^\beta\iota_\beta)\psi$ and induction on the grading degree.  Thus $\hat\Lambda\gamma_\alpha\hat\Lambda^{-1}\psi=\hat\Lambda({\bf e}_\alpha\wedge\hat\Lambda^{-1}\psi-\iota_\alpha\hat\Lambda^{-1}\psi)=\Lambda_\alpha^\beta\gamma_\beta\psi$.
\end{proof}

\begin{theorem}
${\rm tr}(\gamma^\alpha\gamma^\beta\hat\Omega_{\alpha\beta})=8R$
\end{theorem}

\begin{proof}
Compute $\gamma^\alpha\gamma^\beta\hat\Omega_{\alpha\beta}\psi=(g^{\alpha\rho}g^{\beta\sigma}-g^{\alpha\sigma}g^{\beta\rho}){\bf e}_\sigma\wedge\iota_\rho\hat\Omega_{\alpha\beta}\psi+A\psi+B\psi$, where $A$ ($B$) increases (decreases) the grading degree by 2.  One then computes the trace of $\gamma^\alpha\gamma^\beta\hat\Omega_{\alpha\beta}$ when restricted to each grading degree by using the fact that $\hat\Omega_{\alpha\beta}$ satisfies the Leibniz rule and $\hat\Omega_{\alpha\beta}{\bf e}_\rho=R_{\rho\alpha\beta}^\epsilon{\bf e}_\epsilon$.  For example, when restricted to degree 2 we compute
\begin{align*}
  &(g^{\alpha\rho}g^{\beta\sigma}-g^{\alpha\sigma}g^{\beta\rho})
   {\bf e}_\sigma\wedge\iota_\rho\hat\Omega_{\alpha\beta}{\bf e}_{\mu\nu}\\
  &\quad\quad
   =(g^{\alpha\rho}g^{\beta\sigma}-g^{\alpha\sigma}g^{\beta\rho})
    (R_{\rho\mu\alpha\beta}{\bf e}_{\sigma\nu}
     -g_{\rho\nu}R_{\mu\alpha\beta}^\epsilon{\bf e}_{\sigma\epsilon})
    -[\mu\leftrightarrow\nu].
\end{align*}
The trace is obtained by replacing ${\bf e}_{\tau\omega}$ with $\tfrac{1}{2}(\delta_\tau^\mu\delta_\omega^\nu-\delta_\tau^\nu\delta_\omega^\mu)$ on the right hand side of this equation.  The result is ${\rm tr}(\gamma^\alpha\gamma^\beta\hat\Omega_{\alpha\beta})=4R$ when restricted to grading degree 2.  Similarly one computes the trace to be $2R$ in degrees 1 and 3, and trivial in degrees 0 and 4.
\end{proof}

\begin{lemma}\label{lem:igdg}
$\hat{g}^{-1}\hat{g}_{,\nu}$ satisfies the Leibniz rule and $\hat{g}^{-1}\hat{g}_{,\nu}{\bf e}_\rho=g^{\alpha\beta}g_{\alpha\rho\nu}{\bf e}_\beta$
\end{lemma}

\begin{proof}
This follows from the definition of the extended metric and the fact that the derivative of the determinant of a $k\times k$ matrix $M$ is given by the sum of the determinants of the matrics $M_i$ ($1\leq i\leq k$), where $M_i$ is obtained by replacing the $i$--th column of $M$ with its derivative.
\end{proof}

\begin{lemma}\label{lem:extcon-int}
$[\hat\Gamma_\beta,\iota_\alpha]=[\hat{g}^{-1}\hat{g}_{,\beta},\iota_\alpha]+\Gamma_{\alpha\beta}^\epsilon\iota_\epsilon$
\end{lemma}

\begin{proof}
Show that $[\hat{g}^{-1}\hat{g}_{,\beta},\iota_\alpha]=-[\partial_\beta,\iota_\alpha]$ using lemma \ref{lem:igdg} and induction on the grading degree.  Now use lemma \ref{lem:con-int}.
\end{proof}

\begin{theorem}
$T^{\alpha\beta\epsilon}=\gamma^\nu T_\nu^{\alpha\beta\epsilon}-\tfrac{1}{2}(g^{\alpha\epsilon}\gamma^\beta+g^{\beta\epsilon}\gamma^\alpha-g^{\alpha\beta}\gamma^\epsilon)$, where $T_\nu^{\alpha\beta\epsilon}$ satisfies the Leibniz rule and $T_\nu^{\alpha\beta\epsilon}{\bf e}_\rho=\tfrac{1}{2}[\delta_\nu^\alpha(g^{\epsilon\mu}\delta_\rho^\beta-g^{\beta\mu}\delta_\rho^\epsilon)+\delta_\nu^\beta(g^{\epsilon\mu}\delta_\rho^\alpha-g^{\alpha\mu}\delta_\rho^\epsilon)]{\bf e}_\mu$
\end{theorem}

\begin{proof}
Use lemmas \ref{lem:igdg} and \ref{lem:extcon-int} to show that
$$(\hat{g}^{-1}\hat{g}_{,\nu}\gamma^\nu-\{\gamma^\nu,\hat\Gamma_\nu\})\psi
  =g^{\nu\theta}g^{\tau\lambda}g_{\tau\theta\nu}{\bf e}_\lambda\wedge\psi
   +\gamma^\nu(\hat{g}^{-1}\hat{g}_{,\nu}-2\hat\Gamma_\nu)\psi
   -g^{\nu\theta}\Gamma_{\theta\nu}^\lambda\gamma_\lambda\psi
$$
The derivative of the first summand on the right hand side of this equation with respect to $g_{\alpha\beta\epsilon}$ is equal to $-(\partial\gamma^\epsilon/\partial g_{\alpha\beta})\psi$.  Thus from equation \eqref{eq:var2}, we have
$$T^{\alpha\beta\epsilon}\psi
  =\frac{\partial}{\partial g_{\alpha\beta\epsilon}}
     [\gamma^\nu(\hat{g}^{-1}\hat{g}_{,\nu}-2\hat\Gamma_\nu)\psi
      -g^{\nu\theta}\Gamma_{\theta\nu}^\lambda\gamma_\lambda\psi].
$$
It remains to show that $T_\nu^{\alpha\beta\epsilon}=\partial(\hat{g}^{-1}\hat{g}_{,\nu}-2\hat\Gamma_\nu)/\partial g_{\alpha\beta\epsilon}$.  From lemma \ref{lem:igdg}, we see that $T_\nu^{\alpha\beta\epsilon}$ satisfies the Leibniz rule.  Use the same lemma to compute $(\hat{g}^{-1}\hat{g}_{,\nu}-2\hat\Gamma_\nu){\bf e}_\rho=g^{\theta\lambda}(g_{\rho\nu\theta}-g_{\theta\nu\rho}){\bf e}_\lambda$.
\end{proof}

\begin{theorem}
${T^{\alpha\beta\epsilon}}_{;\beta}
  \doteq
   {T^{\alpha\beta\epsilon}}_{,\beta}
   +[\hat\Gamma_\beta,T^{\alpha\beta\epsilon}]
   +\Gamma_{\lambda\beta}^\alpha T^{\lambda\beta\epsilon}
   +\Gamma_{\lambda\beta}^\beta T^{\alpha\lambda\epsilon}
   +\Gamma_{\lambda\beta}^\epsilon T^{\alpha\beta\lambda}
  =0$.
\end{theorem}

\begin{proof}
As per equation \eqref{eq:Tmatrix}, we can write $T^{\alpha\beta\epsilon}=\gamma^\nu T_\nu^{\alpha\beta\epsilon}+S^{\alpha\beta\epsilon}$.  Using equation \eqref{eq:gamma-xconn}, one computes that $S^{\alpha\beta\epsilon}$ is covariantly constant: ${S^{\alpha\beta\epsilon}}_{;\eta}=0$.  Although $T^{\alpha\beta\epsilon}_\nu$ is not covariantly constant, it is divergence--free.  That is,
$${T^{\alpha\beta\epsilon}_\nu}_{;\beta}
  \doteq
  {T^{\alpha\beta\epsilon}_\nu}_{,\beta}
   +[\hat\Gamma_\beta,T^{\alpha\beta\epsilon}_\nu]
   +\Gamma_{\lambda\beta}^\alpha T^{\lambda\beta\epsilon}_\nu
   +\Gamma_{\lambda\beta}^\beta T^{\alpha\lambda\epsilon}_\nu
   +\Gamma_{\lambda\beta}^\epsilon T^{\alpha\beta\lambda}_\nu
   -\Gamma_{\nu\beta}^\lambda T^{\alpha\beta\epsilon}_\lambda
  =0
$$
which is established by noting that ${T^{\alpha\beta\epsilon}_\nu}_{;\beta}$ satifies the Leibniz rule, and computing ${T^{\alpha\beta\epsilon}_\nu}_{;\beta}{\bf e}_\rho=0$.  Equation \eqref{eq:gamma-xconn} again can then be used to show that ${T^{\alpha\beta\epsilon}}_{;\beta}=\gamma^\nu{T^{\alpha\beta\epsilon}_\nu}_{;\beta}+{S^{\alpha\beta\epsilon}}_{;\beta}=0$.
\end{proof}

\begin{theorem}
If $[\gamma^\mu,F_{\alpha\beta}]=0$, then ${\rm tr}_2(\gamma^\alpha\gamma^\beta F_{\alpha\beta})={\rm tr}(F_{\alpha\beta}F^{\alpha\beta})$
\end{theorem}

\begin{proof}
First use the properties of the trace and the Clifford algebra identity for gamma matrices to show that ${\rm tr}(\gamma^\alpha\gamma^\beta F_{\alpha\beta})^2={\rm tr}(\gamma^\alpha\gamma^\beta\gamma^\mu\gamma^\nu F_{\alpha\beta}F_{\mu\nu})=-2\,{\rm tr}(F_{\alpha\beta}F^{\alpha\beta})$.  Now use the matrix identity ${\rm tr}_2(M)=\tfrac{1}{2}({\rm tr}^2M-{\rm tr}M^2)$.
\end{proof}


\end{document}